\documentclass[preprint,showpacs,preprintnumbers,amsmath,amssymb]{revtex4}

\usepackage{graphicx}
\usepackage{dcolumn}
\usepackage{bm}

\newcommand{\w}[1]{\mbox{\boldmath{$#1$}}}

\begin{document}


\title{Gibbsian theory of power law distributions }

\author{R. A. Treumann$^\ddag$\footnote{Visiting the Technical University of Braunschweig} and C. H. Jaroschek$^{**}$
}\email{treumann@issibern.ch}
\affiliation{$^\ddag$ Department of Geophysics and Environmental Sciences, Munich University, D-80333 Munich, Germany  \\ 
$^{**}$Department Earth \& Planetary Science, University of Tokyo, Tokyo, Japan
}%

\date{\today}

\begin{abstract}
It is shown that power law phase space distributions describe marginally stable Gibbsian equilibria far from thermal equilibrium which are expected to occur in collisionless plasmas containing fully developed quasi-stationary turbulence. Gibbsian theory is extended on the fundamental level to statistically dependent subsystems introducing an `ordering parameter' $\kappa$. Particular forms for the entropy and partition functions are derived with super-additive (non-extensive) entropy, and a redefinition of temperature in such systems is given. 
\end{abstract}

\maketitle

Power law phase space distributions of charged particles are at the heart of collisionless plasma physics. In space they have been observed almost everywhere \cite{Vasyliunas1968, Scudder1979,Christon1988,Onsager1991,Lin1996,Fisk2006} posing the problem of reproducing their regular occurrence. Formation of power laws has first been made plausible by Fermi \cite{Fermi1949} in shock acceleration which, however, must be pushed to its margins in order to explain the commonality of  power law distributions under conditions when the plasma is fairly quiet as, for instance, in the solar wind \cite{Fisk2006}. They represent a general property of slowly evolving quasi-stationary collisionless plasmas (or other systems) far from collisional equilibrium but near marginal stability. This raises the question \cite{Treumann1999} whether those systems cannot be described by a version of thermodynamic quasi-equilibrium.  Fisk and Gloeckler \cite{Fisk2006}  recently provided an important  thermodynamic argument for formation of an asymptotic canonical power law tail on the distribution $f(v)\propto v^{-5}$ from turbulent energy cascading with heat flow suppressed. A general physical argument \cite{Treumann1999} was based  on the assumption that strongly turbulent interactions in collisionless plasma require the Boltzmann collision integral to be modified for which an {\it ad hoc} form was proposed. This led to the formulation of a generalized Lorentzian statistical mechanics yielding an equilibrium distribution with power law tail resembling the observed \cite{Vasyliunas1968,Christon1988,Onsager1991,Lin1996} $\kappa$-distributions.  

More specialized mechanisms based on wave particle interaction models \cite{Hasegawa1985} or combinations of inhomogeneity, plasma flow, radiative transport and residual Coulomb interactions \cite{Scudder1979,Collier1993} also yield power law tails. The approach of Hasegawa et al. \cite{Hasegawa1985}  is particularly illuminating. It yields an electron  distribution from nonlinear interaction of an electron plasma with a photon bath, with power $\kappa$ being a function of the self-consistent light wave intensity, applicable to either laser-plasma interaction or high radiation power astrophysical objects. 
Generalizations of thermodynamics based on mathematical {\it ad hoc} modifications of the Boltzmann-Shannon entropy have been in use for several decades \cite{Renyi1970}. Among them the R\'enyi entropy enjoys application in chaos theory, while Tsallis' entropy, a variant of a parameterized version proposed by Daroczy \cite{Renyi1970}, lies at the basis of  a ``non-extensive"  thermodynamics. 
Here, using Gibbsian theory, we show that power law distributions and generalized Lorentzian thermodynamics under closed homogeneous conditions far from equilibrium may arise as the  consequence of the violation of the statistical independence of subsystems.
 
The starting point in statistical equilibrium mechanics is the Gibbs distribution $w_i(\epsilon_i)=A\,\exp\,(-\epsilon_i/T)$, the probability of finding a particle in energy state $\epsilon_i$ in the phase space volume $\Gamma'(\epsilon')$ embedded in a large system of temperature (in energy units)  $T$ \cite{Huang1987}. It arises from the phase space integral $w_i\propto\int {\rm d}\Gamma'\,\delta(\epsilon_i+\epsilon'-E)$, where $E$ is the average total energy of the system, by replacing d$\Gamma'$/d$\epsilon'=\exp[S'(\epsilon')]/\Delta\epsilon'$ with entropy $S'(\epsilon')$. The exponential dependence on entropy implies that the subsystems are uncorrelated. Phase space elements multiply, and $S'$ is additive, i.e. an extensive quantity. This independence breaks down in collisionless plasma turbulence due to the existence of phase space attractors; and $S'$ should loose its additive character. In order to maintain the general argument we seek for a generalization of the Gibbs distribution, i.e. for properly replacing the exponential dependence of the phase space element on $S'$ with another function that in the limit of independence of the subsystems reproduces the exponential. Among the many functions and distributions serving these needs, the simplest {\it real function} with the desired property is the generalized Lorentzian with arbitrary constant (independent of $\epsilon'$) `ordering parameter' $\kappa$,
\begin{equation}
\frac{{\rm d}\Gamma'(\epsilon')}{{\rm d}\epsilon'}=\frac{1}{\Delta\epsilon'}\left[S'_\kappa(E)-\frac{S_\kappa'(\epsilon')}{\kappa}\right]^{-(\kappa +1)}.
\end{equation}
It reproduces Gibbs' phase space element for $\kappa\to\infty$. In addition  $\lim_{\kappa\to\infty}S'_\kappa= S'$, a condition made use of later.  Advantage has also been taken of the freedom that for $\kappa\to\infty$ an arbitrary constant (taken here as 1) can be added to the power $\kappa$ without changing the result. The identity with Gibbs' expression  in the limiting case is proved by taking logarithms. In the above expression $S'_\kappa(E)$ is the constant total entropy which must be added  in order to avoid an `infrared catastrophe' at $S'_\kappa(\epsilon')\to 0$.

Expanding the entropy in the integral around its value at total energy $E$ with respect to energy in state $i$ yields $S_\kappa'(E-\epsilon_i)\simeq S_\kappa'(E) - \epsilon_i [{\rm d}S_\kappa'(E)/{\rm d}E]= S_\kappa'(E) - \epsilon_i /T'$, with temperature $1/T'= {\rm d}S_\kappa'(E)/{\rm d}E$. Solving it yields the probability distribution in the canonical ensemble 
\begin{equation}
w_{i,\kappa}(\epsilon_i) = A'\left[S'(E)\left(1-\frac{1}{\kappa}\right) +\frac{\epsilon_i}{\kappa T'}\right]^{\!-(\kappa +1)}.
\end{equation}
The first term in the brackets can be absorbed into temperature and normalization constant yielding
\begin{equation}
w_{i,\kappa}(\epsilon_i) = A\left(1 +\frac{\epsilon_i}{\kappa T_\kappa}\right)^{\!\!-(\kappa +1)},
\end{equation}
with new subsystem temperature $T_\kappa=S_\kappa'(E)(1\!-\!1/\kappa)(\partial S_\kappa'/\partial E)^{-1}$ depending  on $S_\kappa'(E)$. The usual definition of the temperature $T$ recovers for $\kappa\to\infty$. Due to this unusual dependence on total entropy, $\kappa$-temperatures $T_\kappa$ turn out quite small, in particular for large total entropy. Knowledge of the `real' temperature $T'$ requires determination of the total entropy $S_\kappa'(E)$. In practical applications this provides a severe complication since the total entropy in this case is implicitly expressed through average energy $E$ from the first thermodynamic law. This is seen from Eq. (11) below, where $S(E)$ appears in differential form in the temperature and also in the normalization integral $A$, yielding an integro-differential equation for $S_\kappa(E)$.

At large $\epsilon_i$ the new canonical probability distribution is a relativistically correct power law distribution.  Replacing $\epsilon_i= p_i^2/2m$ non-relativistically [or relativistically $\epsilon_i=mc^2\gamma({\w p}_i)$] with particle momentum ${\w p_i}$, it becomes power law in ${\w p_i}$.  It must be normalized to one summing over all states. This yields the normalization constant $A$
\begin{equation}
\sum_i w_{i,\kappa} = A \sum_i \left[1 +\frac{\epsilon_i}{\kappa T_\kappa}\right]^{-(\kappa +1)} \!\!\!= 1.
\end{equation}

It is straightforward to go from discrete probability distributions to classical distribution functions by defining  a continuous  phase space distribution $f_\kappa({\w p, \w x})$ with energy variable $\epsilon ({\w p, \w x})$ continuous in momentum ${\w p}$ and space ${\w x}$,
\begin{equation}
f_\kappa({\w p, \w x}) = A \left[1 +\frac{\epsilon ({\w p, \w x})}{\kappa T_\kappa}\right]^{-(\kappa +1)},
\end{equation}
and normalization $\int f_\kappa({\w p, x}) {\rm d}{\w p}\,{\rm d}{\w x}=1$. In the classical case of constant particle number $N$ and volume $V$ it replaces the Maxwell distribution [a more conventional choice is  normalization to number density $N/V$, which requires introducing the $s$-dimensional phase space element $(2\pi\hbar)^s$] .

Distributions of this kind have been applied in space plasma physics in various approximations. At high particle energies $\epsilon/\kappa T_\kappa\gg 1$, $f_\kappa$ becomes a simple power law distribution $f_\kappa(\epsilon)\propto \epsilon^{-(1+\kappa)}$. Its second moment gives the average energy density ${\cal E}$ of the system
\begin{equation}
{\cal E} = A \int\epsilon^\frac{3}{2} \left[1 +\frac{\epsilon ({\w p, \w x})}{\kappa T_\kappa}\right]^{-(\kappa +1)}{\rm d}\epsilon. 
\end{equation}
For reasons of convergence the lower bound $\kappa_{\rm min}\geq 3/2$ is set on the power law index $\kappa$, translating to marginal flattest (non-relativistic) power law distributions  $f_\kappa(\epsilon) \propto\epsilon^{-5/2}$ in energy  or $f_\kappa(\w p) \propto p^{-5}$ in momentum. This theoretical power is in agreement with observation and nicely confirms the thermodynamic arguments of Fisk and Gloeckler \cite{Fisk2006} in the absence of heat flux. Nonzero heat flux requires the existence of the next higher moment  implying $f_\kappa(\w p)\propto p^{-7}$. The general condition on the value of $\kappa$ for the $l$th moment ${\textsf M}_{l}\propto\int {\w p}^lf_\kappa{\rm d^3}p$ to exist $(l=0,1,2,\dots)$ is $\kappa\geq\frac{1}{2}(l+1)$. Power law particle distributions observed in collisionless space plasmas  frequently exhibit steeper slopes $3/2<\kappa\lesssim 10$ \cite{Christon1988}, suggesting either presence of higher moments of $f_\kappa$ or particle losses through real space boundaries. At relativistic energies $\epsilon\simeq pc$ the asymptotic behavior of the distribution function changes to $f_\kappa(\epsilon)\propto \epsilon^{-4}$, and the energy distribution exhibits a break at  $\epsilon\gg mc^2$ changing from power -2.5 to power -4. 

So far we dealt with constant particle number $N$. It is, however, simple matter to take into account variations in $N$  in the same way as in Gibbsian statistics, letting the infinitesimal phase space volume ${\rm d}\Gamma'$ and entropy $S_\kappa'(N)$ depend on $N$, with $N_0$  the total particle number. Then the delta function in the integral defining $w_i$ also depends on $N$ according to $\delta(\epsilon_i+\epsilon'-E, N+N'-N_0)$, and the integration is with respect to the primed coordinates. Keeping the volume constant, the entropy is now expanded with respect to energy and particle number yielding $S_\kappa'(E-\epsilon_{i,N},N_0-N )\simeq S_\kappa'(E,N_0)  - \epsilon_i /T' +\mu N/T'$. The factor in front of $N$ is the chemical potential $\mu = -T'(\partial S_\kappa'/\partial N_0)_{E,V}$. Redefining the temperature as done above, the probability distribution function of the grand canonical ensemble becomes
\begin{equation}
w_{i,\kappa}(\epsilon_i, N) =A\left[1 +\frac{(\epsilon_i-\mu N)}{\kappa T_\kappa}\right]^{-(\kappa +1)}.
\end{equation}
The normalization condition now takes into account the summation over particle numbers in the different states,
\begin{equation}
\sum\limits_{N,i} w_{i,\kappa}(\epsilon_i, N)=A\sum\limits_{N,i} \left[1 +\frac{(\epsilon_i-\mu N)}{\kappa T_\kappa}\right]^{\!-(\kappa +1)}\!\!\!\!\! = 1,
\end{equation}
first summing over energy states $i$ at constant $N$, and then summing over all $N$. This distribution function cannot as easily as before be transformed into a classical phase space distribution function. If, for the moment, we restrict to fixed $N$, the distribution in the classical limit takes the differential form d$w(N)= f({\w p_N, \w x_N}) {\rm d}{\w p_N}{\rm d}{\w x_N}$, and the phase space distribution can be determined for the $N$th subspace $( {\w p_N},{\w x_N})$ of phase space. It requires knowledge of the analytical form of the quantity underlying the whole theory, viz. the entropy $S_\kappa$.

In Gibbs-Boltzmann statistical mechanics the entropy is the ensemble average over the logarithm of the probability distribution $S=-\langle \log w_{i}(N)\rangle$. The angular brackets stand for the ensemble average $\sum_{i,N}w_{i}(N)\log w_{i}(N)$. Inserting the above $\kappa$-probability distribution function one realizes the impossibility to obtain the basic thermodynamic relations from this definition. A modified definition of entropy is needed where $S_\kappa$ is given as the ensemble average of the logarithm of a functional $g[w_{i,\kappa}(N)]$ of the probability distribution as
\begin{equation}
S_\kappa=-\langle \log g[w_{i,\kappa}(N)]\rangle
\end{equation}
This functional must be chosen such that in the limit $\kappa\to\infty$ it reproduces the grand canonical distribution $w_{i,\kappa}(N)$. With the grand canonical probability distribution the only possible choice for $g$ is 
\begin{equation}
g[w_{i,\kappa}(N)] = A \exp\left\{\kappa\left[1-\left(\frac{A}{w_{i,\kappa}(N)}\right)^{\!\!\frac{1}{1+\kappa}}\right]\right\}
\end{equation}
with the same normalization constant $A$ as in $w_{i,\kappa}$, as it is easy to show that $g\to w_{i,\kappa}(N)$ for $\kappa\to\infty$.   Inserting into $S_\kappa$, with average energy $\langle\epsilon_{iN}\rangle=E$, one finds that
\begin{equation}
S_\kappa = -\log A + E /T_\kappa - \mu \langle N\rangle/T_\kappa.
\end{equation}
Rearranging just reproduces Gibbs' grand canonical thermodynamic potential $\Omega_\kappa = F_\kappa- \mu\langle N\rangle = E -T_\kappa S_\kappa-\mu\langle N\rangle\equiv T_\kappa\log A$ or, when using the normalization condition, yields expressions for the free energy $F_\kappa$ and $\Omega_\kappa$, the latter being
\begin{equation}
\Omega_\kappa = -T_\kappa\log \sum\limits_{i,N} \left(1+\frac{\epsilon_{iN}-\mu N}{\kappa T_\kappa}\right)^{\!\!-(1+\kappa)}
\end{equation}
From this formula the grand partition function follows as the sum of the $N$ canonical partition functions $Z_{\kappa,N}$
\begin{equation}
Z_\kappa \equiv \frac{1}{A}=\sum\limits_N Z_{\kappa, N}=\sum\limits_{i,N} \left(1+\frac{\epsilon_{iN}-\mu N}{\kappa T_\kappa}\right)^{\!\!-(1+\kappa)}
\end{equation}
From it follow all thermodynamic and statistical mechanical quantities of a $\kappa$ plasma. 

From $\Omega_\kappa=T_\kappa\log A$  it is straightforward to write down the grand canonical probability distribution function 
\begin{equation}
w_{i,\kappa}(\epsilon_i, N) =\left[1 +\frac{(\epsilon_{iN}-\mu N)}{\kappa T_\kappa}\right]^{\!-(\kappa +1)}\!\!\!\!{\rm e}^{\Omega_\kappa/T_\kappa},
\end{equation}
showing that its dependence on $\Omega_\kappa$ in $\kappa$-theory is the same as in ordinary Gibbsian theory. This allows to write down the phase space distribution function in $N$th subspace as
\begin{equation}
f_\kappa({\w p_N, \w x_N})\! =\! \frac{{\rm e}^{\Omega_\kappa/T_\kappa}}{(2\pi\hbar)^{s}}\!\left[1 +\!\frac{\epsilon({\w p_N,\w x_N})-\mu N}{\kappa T_\kappa}\right]^{\!-(\kappa +1)}
\end{equation}
where $s$ is the dimensionality of the $N$th subspace. Again, the energy distribution remains to be power law. However, now the distribution also depends on particle number $N$ and  chemical potential $\mu$ which is determined through normalizing to particle number density $\langle N\rangle/V$ when summing up all contributions from the subspaces. This is a formidable task that cannot be completed without precise knowledge of the energy states, i.e. $\epsilon({\w p_N,\w x_N})$. 

An approximate expression can, however, be obtained assuming that the mean number of particles in each subspace is very small. In this case we may write for the mean one particle distribution function
\begin{equation}
{\bar f}_{\kappa,1}({\w p_1}) \propto \left[1 +\frac{\epsilon({\w  p_1})-\mu }{\kappa T_\kappa}\right]^{-(\kappa +1)}
\end{equation}
which is the generalization of the Boltzmann distribution. Here we have put $N=1$ for the occupied elements of phase space, retained only momentum dependence and absorbed the space dependence (giving a volume factor $V$) into the factor in front. The dependence on chemical potential $\mu$ is again retained and is determined through number density $\langle N\rangle/V$. [In application to measurements the factor $1-\mu/\kappa T_\kappa$ can be taken out, leading to another re-definition of temperature.]

We briefly discuss in passing two simple cases that can be constructed from $\Omega_\kappa$, the cases of Fermi and Bose distributions. We write $\Omega_\kappa$ in terms of occupation numbers $n_i$ of the $i$th energy level instead of $N$
\begin{equation}
\Omega_\kappa = -T_\kappa\log \sum\limits_{n_i} \left[1+n_i\frac{\epsilon_i-\mu}{\kappa T_\kappa}\right]^{-(1+\kappa)}.
\end{equation}
For the Fermi case $n_i=0,1$ this becomes
\begin{equation}
\Omega_{\kappa,F} = -T_\kappa\log \left\{1+ \left[1+\frac{\epsilon_i-\mu}{\kappa T_\kappa}\right]^{-(1+\kappa)}\right\}.
\end{equation}
Calculating the average occupation number ${\bar n_{i,F}}= - \partial\Omega_{\kappa,F}/\partial\mu$ yields the correct Fermi $\kappa$-distribution
\begin{equation}
{\bar n_{i,F}} =\left(1+\frac{1}{\kappa}\right)\frac{[1+(\epsilon_i-\mu)/\kappa T_\kappa]^{-1}}{\{1+[1+(\epsilon_i-\mu)/\kappa T_\kappa]^{1+\kappa}\}}.
\end{equation}
It has no zero temperature limit except for $\kappa\to\infty$ where it becomes the usual Fermi function. The only possible solution with positive chemical potential would be a condensation of all particles at one energy level $\epsilon_i=\mu$, which can be determined from the energy integral but is strictly forbidden by the Pauli exclusion principle. (Anyonic occupations would be possible, however.) It thus exclusively describes finite temperature states, implying only negative chemical potentials $\mu<-\kappa T_\kappa$ and thus no degeneration. This is reasonable as correlations should occur only at finite high enough temperature.  

For the Bose distribution we sum over all $n_i=0, 1, \dots$ 
\begin{equation}
{\bar n_{i,B}} \!=\!\left(1+\frac{1}{\kappa}\right)\! \frac{\sum_{n_i}n_i[1+n_i(\epsilon_i-\mu)/\kappa T_\kappa]^{-(2+\kappa)}}{\sum_{n_i}[1+n_i(\epsilon_i-\mu)/\kappa T_\kappa]^{-(1+\kappa)}}
\end{equation}
There is no way of bringing this into closed form.  Clearly, $\mu\leq 0$. Similar to the Fermi case there is no zero-temperature limit and thus also no condensation. One thus concludes that the Bose $\kappa$-distribution is as well defined only for finite $T$, which again is reasonable as it is defined for correlated states evolving at finite temperature only. 

The present approach is valid for closed homogeneous turbulent systems. It treats $\kappa$ as an {\it ad hoc} parameter containing the hidden correlations. It is determined from observation and is a function of the power in the turbulent field fluctuations in stationary turbulent quasi-equilibrium in the absence of binary collisions. Its functional dependence requires solving the complete wave-particle dynamics, as was done only for special cases like in the work of Hasegawa et al. \cite{Hasegawa1985} for electrons interacting with a photon bath. Scudder and Olbert \cite{Scudder1979} include plasma flow, rudimentary Coulomb collisions, inhomogeneity, non-locality and radiation transport in Boltzmann theory to construct power law distributions. The thermodynamic approach of Fisk and Gloeckler \cite{Fisk2006}  resembles ours in determining the marginal $\kappa$. In our spirit $\kappa$ evolves slowly until time approaches the binary collision time, when it starts diverging explosively and Boltzmann statistical mechanics takes over. For $t<\nu_c^{-1}$, $\kappa(t)=\kappa_{min}\left[1-(\nu_c t)^r\right]^{-1}$ can be modeled as an explosive function of time $t$ and Coulomb collision time $\nu_c^{-1}=16\pi N_e\lambda_D^3/\omega_{pe}$.
The exponent $r>1$ controls the strength of the transition from turbulent to collisional equilibrium [$N_e,\omega_{pe},\lambda_D$ are electron density, plasma frequency, and Debye length, respectively]. For spatial applications it is more convenient to write $\kappa(t)$  in terms of  distance $L$ and collisional mean free path $\lambda_{\rm mfp}=v_e/\omega_{pe}$ as $\kappa(L)=\kappa_{min}\left[1-(L/\lambda_{\rm mfp})^r\right]^{-1}$. Deviations of $\kappa$ from $\kappa_{min}$ are measures of its evolution. 


We have constructed a statistical mechanical theory of power law distributions via generalizing Gibbsian theory, relaxing the assumption of independence of subsystems through introducing a generalized Lorentzian form as the {\it simplest real-function generalization of the Gibbs function dependence on entropy}. It contains an {\it ad hoc} ordering parameter $\kappa$ that controls the strength of the subsystem correlations. Particular forms of entropy and partition function have been obtained uniquely as the only means of satisfying the fundamental thermodynamic relations. As classical equilibrium distribution the $\kappa$-distribution was recovered. It replaces the Boltzmann distribution in correlated collisionless quasi-equilibria. This theory extends classical statistical mechanics to correlated collisionless systems where subsystems are not anymore statistically independent. In such systems the entropy is not additive; it is  super-additive (or super-extensive) because the interdependence of subsystems contributes an extra amount to entropy. This can be shown by direct calculation of the total entropy $S_{1+2}\geq S_1+S_2$ of two subsystems $S_1, S_2$. Moreover, a H-theorem $H(t)\equiv\int {\rm d}\Gamma f\ln g[f]$ holds. Taking the time derivative yields ${\rm d}H/{\rm d}t=\int {\rm d}\Gamma \left\{\ln g+{\delta g/g\delta f}\right\}({\rm d}f/{\rm d}t)\leq 0$ as the term in the braces is always positive. For small deviations from equilibrium and $\partial f/\partial t<0$ the system monotonically returns to equilibrium with ${\rm d}f/{\rm d}t=0$. 

Basing the Gibbs-Lorentzian function on counting statistics as in conventional statistical mechanics is not in sight. Counting states in equivalence to throwing dices implies statistical independence. Breaking independence requires  prescription of a particular form of interdependence of subsystems. Thus any counting, if at all possible, must be model-dependent.


{\small
{\it Acknowledgements}. 
We acknowledge discussions with A. Balogh, L. Fisk, G. Gloeckler, K.-H. Glassmeier, M. Hoshino, R. Jokipii,  and H.-R. Mueller, as also the Referees' constructive comments. 
Research supported by the German Ministerium f\"ur Wirtschaft und Technologie and Deutsches Zentrum f\"ur Luft- und Raumfahrt under grant 50OC0103.}






\begin{thebibliography}{}


\bibitem[\protect\citeauthoryear{Vasyliunas}{1968}]{Vasyliunas1968}
Vasyliunas V. M.,  {\it J. Geophys. Res.} {\bf 73}, 2839 (1968).

\bibitem[\protect\citeauthoryear{Christon et al.}{1988}]{Christon1988}
Christon S. et al., {\it J. Geophys. Res.} {\bf 93}, 2562 (1988);  {\bf 96}, 1 (1991).

\bibitem[\protect\citeauthoryear{Onsager et al.}{1991}]{Onsager1991}
Onsager T. G. et al., {\it J. Geophys. Res.} {\bf 96}, 20999 (1991); Paschalidis N. P. et al., {\bf 99}, 8687 (1994).

\bibitem[\protect\citeauthoryear{Lin et al.}{1996}]{Lin1996}
Lin R. P. et al., {\it Geophys. Res. Lett.} {\bf 23}, 1211 (1996).



\bibitem[\protect\citeauthoryear{Scudder and Olbert}{1979}]{Scudder1979}
Scudder J. D. and Olbert S.,  {\it  J. Geophys. Res.} {\bf 84}, 2755 and 6603 (1979).





\bibitem[\protect\citeauthoryear{Fisk and Gloeckler}{2006}]{Fisk2006}
Fisk L. and Gloeckler G., {\it Astrophys. J.}  {\bf 640}, L79 (2006); {\it Proc. NAS}  {\bf 104}, 5749 (2007); 
Gloeckler G. and Fisk L., {\it Astrophys. J.}  {\bf 648}, L63 (2006).


\bibitem[\protect\citeauthoryear{Fermi}{1949}]{Fermi1949}
Fermi E., {\it Phys. Rev.} {\bf 75}, 1169 (1949).

\bibitem[\protect\citeauthoryear{Treumann}{1999}]{Treumann1999}
Treumann R. A.,  {\it Phys. Scripta} {\bf 59}, 19 and 204 (1999); Treumann R. A. et al., {\it Phys. Plasmas} {\bf 11}, 1317 (2004).


\bibitem[\protect\citeauthoryear{Hasegawa et al.}{1985}]{Hasegawa1985}
Hasegawa A., Mima K. and Duong-van M.,  {\it Phys. Rev. Lett.} {\bf 54}, 2608 (1985).


\bibitem[\protect\citeauthoryear{Collier}{1993}]{Collier1993}
Collier M. R., {\it Geophys. Res. Lett.} {\bf 20}, 1531 (1993); Ma C.-Y. and Summers D., {\it Geophys. Res. Lett.} {\bf 25}, 4099 (1998).





\bibitem[\protect\citeauthoryear{R\'enyi}{1970}]{Renyi1970}
Balatoni J. and R\'enyi A., {\it Publ. Math. Inst. Hungar. Acad. Sci.} {\bf 1}, 9 (1956); R\'enyi A., {\it Probability Theory} (North-Holland, Amsterdam, 1970); Daroczy Z., {\it Information Control} {\bf 16}, 36 (1970); Wehrl A., {\it Rev. Mod. Phys.} {\bf 50}, 221 (1978); Tsallis C., {\it J. Stat. Phys.} {\bf 52}, 479 (1988).







\bibitem[\protect\citeauthoryear{Huang}{1987}]{Huang1987}
Huang K.,  {\it Statistical Mechanics}, (John Wiley \& Sons, New York 1987); Landau L. D. and Lifschitz  E. M.,  {\it Statistical Physics}, Vol. 5 (Butterworth-Heinemann, Oxford 1997).











\end{thebibliography}


\end{document}